\documentclass[showpacs,amsmath,twocolumn,showkeys]{revtex4}
\usepackage{dcolumn}    
\usepackage{bm}         
\usepackage{ifpdf}
\usepackage{amssymb,lineno,amsfonts}
\usepackage{graphicx}   
\usepackage{dcolumn}    
\usepackage{bm}         
\usepackage{bbm}
\usepackage{mathrsfs}
\usepackage{upgreek}
\usepackage{mathtools}
\usepackage{epstopdf}
\usepackage{setspace}
\usepackage{hyperref}

\begin{document}

\title{Evolution of Quantum Discord and its Stability in Two-Qubit NMR Systems}

\author{Hemant Katiyar, Soumya Singha Roy, T.S. Mahesh}
\email{mahesh.ts@iiserpune.ac.in}
\affiliation{Department of Physics and NMR Research Center,\\
Indian Institute of Science Education and Research, Pune 411008, India}
\author{Apoorva Patel}
\email{adpatel@cts.iisc.ernet.in}
\affiliation{CHEP and SERC,
Indian Institute of Science, Bangalore 560012, India}

\begin{abstract}
We investigate evolution of quantum correlations in ensembles of two-qubit
nuclear spin systems via nuclear magnetic resonance techniques. We use
discord as a measure of quantum correlations and the Werner state as an
explicit example. We first introduce different ways of measuring discord
and geometric discord in two-qubit systems and then describe the following
experimental studies:
(a) We quantitatively measure discord for Werner-like states prepared using
an entangling pulse sequence. An initial thermal state with zero discord is
gradually and periodically transformed into a mixed state with maximum discord.
The experimental and simulated behavior of rise and fall of discord agree
fairly well.
(b) We examine the efficiency of dynamical decoupling sequences in preserving
quantum correlations. In our experimental setup, the dynamical decoupling
sequences preserved the traceless parts of the density matrices at high
fidelity. But they could not maintain the purity of the quantum states and so were unable to keep the discord from decaying.
(c) We observe the evolution of discord for a singlet-triplet mixed state
during a radio-frequency spin-lock. A simple relaxation model describes
the evolution of discord, and the accompanying evolution of fidelity of
the long-lived singlet state, reasonably well.
\end{abstract}

\maketitle

\section{Introduction}

Since its introduction by Schr\"odinger, entanglement has remained an
extensively studied and yet a mysterious aspect of quantum theory.
Entanglement appears as a by-product of the quantum formalism that assigns
probability amplitudes to physical states and lets them exist in coherent
superpositions. Although, it runs counter to the human intuition gained
through experiences with classical systems, experimental evidence has
consistently favored the existence of such superposed quantum states.
Entanglement was thought to be an indispensable resource for quantum
information processing, which can outperform the corresponding classical
information processing. Indeed, various quantum algorithms that exploit
entanglement have been proposed and successfully tested \cite{horodecki}.
Separable (i.e. not entangled) states were considered insufficient to
implement quantum information processing. That belief has changed since
Ollivier and Zurek \cite{Ollivier} as well as Henderson and Vedral
\cite{vedral} independently introduced a new measure of non-classical
correlations named `\textit{discord}'. Discord is based on the measure of mutual
information between two parts of a system. It can be put in one-to-one
correspondence with entanglement for pure states, but, unlike entanglement,
it can be nonzero for separable mixed states \cite{vedral}.

There have been long debates on the necessity of entanglement for quantum
information processing. For example, the purity of typical spin systems
used in nuclear magnetic resonance (NMR) experiments is too small to exhibit
entanglement; nevertheless, NMR quantum information processing is considered
an efficient test bed for quantum algorithms \cite{Braun,oliviera,chuang}.
Furthermore, Knill and Laflamme proposed an algorithm, called DQC1, which
estimates the trace of any unitary matrix faster than any known classical
algorithm \cite{Knill}. Datta {\it et al.} showed that entanglement in the
DQC1 algorithm is vanishingly small, and it further decays with increase in
the number of qubits \cite{Animesh}. They also showed that the DQC1 algorithm
involves nonzero discord. Thus, our present notion of quantum speed up is tied
to discord rather than entanglement \cite{vedral_prx}. For a more detailed
review on discord, see Ref. \cite{vedral_revarticle}.

Discord happens to be one of the many quantities that can measure the
nonclassicality of a given quantum system. Its standard definition equates
discord to the difference between two classically equivalent forms of mutual
information. Due to the difficulty in measuring this difference, Dakic
{\it et al.} proposed an alternate quantity called `\textit{geometric discord}'.
It is the distance between the given quantum state and the closest classical
state and is easier to quantify than discord \cite{Dakic}. We look at both
of these quantities in our study.

Several experiments to demonstrate quantum correlations in liquid-state NMR
systems have been performed in recent years, e.g., by measuring a suitable
witness operator \cite{serraaug11}, by measuring discord \cite{laflammeoct11},
and by evaluating the Leggett-Garg inequality \cite{soumyaprl}. Measurements
of discord in mixed states have also been performed using optical systems
\cite{white08} and quadrupolar NMR systems \cite{serradiscordquad}.
Here we report an experimental study of time evolution of discord in NMR
systems. After preparing a two-qubit Werner state, we study the accumulation
of discord, effects of dynamical decoupling sequences on it, and its decay
due to decoherence. In Sec. II, we revisit some theoretical aspects of
discord and describe different ways of measuring it for the Werner state.
Then, in Sec. III, we present the experimental details and discuss the
results. We conclude in Sec. IV with some inferences from our analysis.

\section{Theory}

\subsection{Discord}
{\it Conditional Entropy: }
In classical information theory the amount of information contained in a
random variable $X$ is quantified as the Shannon entropy,
\begin{equation}
{\mathcal H}(X)=-\sum_{x}p_{x}\log_{2}p_{x} ~,
\end{equation}
where $p_{x}$ is the probability of occurrence of event $X$. When ${\mathcal H}(X)=0$,
the random variable $X$ is completely determined and no new information is
gained by measuring it. Hence, Shannon entropy can be interpreted as either
the uncertainty before measuring a random variable or the information gained
on measuring it.

Consider a bipartite system containing two subsystems (or random variables),
A and B. Conditional entropy of B quantifies the uncertainty in measurement
of B when A is known, and is represented by ${\mathcal H}(B|A)$. Using classical
probability theory, it can be expressed as 
\begin{equation}
{\mathcal H}(B|A)={\mathcal H}(A,B)-{\mathcal H}(A) ~,
\label{hba}
\end{equation}
where ${\mathcal H}(A,B)$ is the information content of the full system and ${\mathcal H}(A)$ is
the information content of the subsystem A. An equivalent way of defining
the conditional entropy is
\begin{equation}
{\mathcal H}(B|A) = \sum_{i}p_{i}^{a} {\mathcal H}(B|a=i) ~,
\label{eq2}
\end{equation}
where
\begin{equation}
{\mathcal H}(B|a=i) = -\sum_{j}p(b_{j}|a_{i})\log_2{p(b_{j}|a_{i})} ~,
\label{eq3}
\end{equation}
and $p(b_{j}|a_{i})$ is the conditional probability of occurrence of event
$b_{j}$ given that event $a_{i}$ has occurred. Unlike the definition in
Eq. (\ref{hba}), the definition in Eq. (\ref{eq2}) involves measurement of
one subsystem of a bipartite system.

\begin{figure}[b]
\includegraphics[width=4cm]{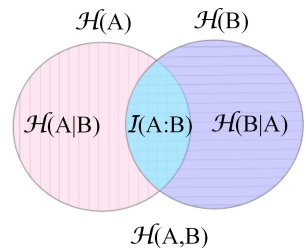}
\caption{The Venn diagram representing total information ${\mathcal H}(A,B)$,
individual informations (${\mathcal H}(A)$, ${\mathcal H}(B)$), conditional
information (${\mathcal H}(A|B)$, ${\mathcal H}(B|A)$), and mutual information
${\mathcal I}(A:B) = {\mathcal J}(A:B)$ in classical information theory.}
\label{mutual}
\end{figure} 

{\it Mutual Information:}
It is the amount of information that is common to both the subsystems of a
bipartite system, and is given by
\begin{equation}
{\mathcal I}(A:B)= {\mathcal H}(A) + {\mathcal H}(B) - {\mathcal H}(A,B) ~.
\label{eq5}
\end{equation}
This expression can be intuitively understood with the help of
Fig.~\ref{mutual}. On the right-hand side, the first two terms quantify the
information content of subsystems A and B, respectively. So the information
common to both the subsystems is counted twice. Subtracting the information
content of the combined system then gives the common (or mutual) information. 
The result is clearly symmetric, i.e., ${\mathcal I}(A:B) = {\mathcal I}(B:A)$. A classically
equivalent form of mutual information, also shown in Fig.~\ref{mutual}, is
\begin{eqnarray}
{\mathcal J}(A:B) &=& {\mathcal H}(B) - {\mathcal H}(B|A) \nonumber \\
&=& {\mathcal H}(B) - \sum_{i}p_{i}^{a}{\mathcal H}(B|a=i) ~,
\label{eq6}
\end{eqnarray}
which removes from the information content of subsystem B the conditional
contribution that is not contained in subsystem A.
 
In quantum information theory, the von Neumann entropy gives the
information content of a density matrix,
\begin{equation}
H(\rho) = -\sum_{x}\lambda_{x}\log_{2}\lambda_{x} ~,
\label{eq7}
\end{equation}
where $\lambda_{x}$ are the eigenvalues of the density matrix $\rho$.
Although the two expressions of mutual information, Eqs.~(\ref{eq5}) and
(\ref{eq6}), are equivalent in classical information theory, they are not
so in quantum information theory. The reason for the difference is that
the expression for mutual information given by Eq.~(\ref{eq5}) involves
measurement and depends on its outcomes. Measurements in quantum theory
are basis dependent and also change the state of the system. Henderson
and Vedral \cite{vedral} have proved that the total classical correlation
can be obtained as the largest value of 
\begin{eqnarray}
J(A:B) &=& H(B) - H(B|A) \nonumber \\
&=& H(B) - \sum_{i}p_{i}^{a}H(B|a=i) ~,
\label{eq8}
\end{eqnarray}
where the maximization is performed over all possible orthonormal measurement
bases $\{\Pi_{i}^{a}\}$ for A, satisfying $\sum_{i}\Pi_{i}^{a}=\mathbbm{1}$
and $\Pi_{i}^{a}\Pi_{j}^{a} = \delta_{ij}\Pi_{i}^{a}$ \cite{footPOVM}.
Therefore, the non-classical correlations can be quantified as the difference
\begin{equation}
D(B|A) = I(A:B) - \max_{\{\Pi_{i}^{a}\}}J(A:B) ~.
\label{disc}
\end{equation}
Ollivier and Zurek named this difference `\textit{discord}' \cite{Ollivier}.
Zero-discord states or ``classical'' states are the states in which the
maximal information about a subsystem can be obtained without disturbing
its correlations with the rest of the system.

Discord is not a symmetric function in general, i.e. $D(B|A)$ and $D(A|B)$
can differ. Datta \cite{animesh_nullity} has proved that a state
$\rho_{AB}$ satisfies $D(B|A)=0$ if and only if there exists a complete
set of orthonormal measurement operators on $A$ such that
\begin{equation}
\rho_{AB} = \sum_{i}p_{i}^{a} \Pi_{i}^{a}\otimes\rho_{B|a=i} ~.
\label{classical}
\end{equation}
When one part of a general bipartite system is measured, the resulting
density matrix is of the form given by Eq.~(\ref{classical}). Since the
state rendered on measurement is a classical state, one can extract
classical correlations from it. Thus, for any quantum state and every
orthonormal measurement basis, there exists a classically correlated state.
Maximization of $J(A:B)$ captures the maximum classical correlation that
can be extracted from the system, and whatever extra correlation that may
remain is the quantum correlation.


\subsection{Evaluation of Discord}
Given a density matrix $\rho_{AB}$, it is easy to construct the reduced density
matrices $\rho_{A}$ and $\rho_{B}$ and then obtain the total correlation
$I(A:B)$ using the quantum analog of Eq.(\ref{eq5}). Maximization of $J(A:B)$
to evaluate discord is nontrivial, however. The brute force method is to
maximize $J(A:B)$ over as many orthonormal measurement bases as possible,
taking into account all constraints and symmetries. For a general quantum
state, a closed analytic formula for discord does not exist, but for certain
special class of states analytical results are available \cite{girolami}. 
For example, Chen {\it et al.} have described analytical evaluation of
discord for two qubit X states under specific circumstances \cite{chen}. 
Luo has given an analytical formula for discord of the Bell-diagonal states
that form a subset of the X states \cite{luo}. In our work, we evaluate
discord using both the brute force method and the Luo method. We use
Bell-diagonal states in our experiments, but our experimental preparation
of the states is not perfect. The difference between the discord values
obtained by the two methods then provides an estimate of the experimental
imperfections.

{\it Extensive measurement method:} 
This method involves measurements over extensive sets of orthonormal basis
vectors and maximization of $J(A:B)$. For measurement of a single qubit in
a two-qubit system, we use the orthonormal basis
\begin{equation}
\{\vert u\rangle = \cos\theta\vert 0\rangle
                 + e^{i\phi}\sin\theta\vert 1\rangle ~,~
  \vert v\rangle = \sin\theta\vert 0\rangle
                 - e^{i\phi}\cos\theta\vert 1\rangle \} ~,
\label{basisvec}
\end{equation}
and let $\cos\theta\in [-1,1]$ and $\phi\in[0,2\pi)$ vary in small steps.
For every choice of $\theta$ and $\phi$, we project the experimental
density matrix obtained by tomography along the orthonormal basis.
The postprojection density matrix is
\begin{eqnarray}
\rho' = \sum_{i=u,v} \Pi_{i}^{a} \rho \Pi_{i}^{a}
      = \sum_{i=u,v} p_{i}^{a} \Pi_{i}^{a}\otimes\rho_{B|a=i} ~,
\end{eqnarray}
with $p_{i}^{a} = {\rm Tr}[\Pi_{i}^{a}\rho]$. Discord is then
obtained from the conditional density matrix $\rho_{B|a=i}$ as
per Eqs.~(\ref{eq8}) and (\ref{disc}).

Strictly speaking, this method gives a lower bound on $J(A:B)$, since the
direction maximizing $J(A:B)$ may not exactly match any of the points on
the discrete $(\theta,\phi)$ grid. Also, when the desired state is
isotropic, e.g., the Werner state, the angular variation of $J(A:B)$
provides an estimate of the inaccuracy in the state preparation, e.g.,
due to inhomogeneities and pulse imperfections.

{\it Analytical method for the Bell-diagonal states:}
As the name suggests, the Bell-diagonal states are diagonal in the Bell
basis, given by 
\begin{equation}
\vert \psi^{\pm}\rangle = \frac{1}{\sqrt{2}}
                        ( \vert 01\rangle \pm \vert 10\rangle ) ~,~~
\vert \phi^{\pm}\rangle = \frac{1}{\sqrt{2}}
                        ( \vert 00\rangle \pm \vert 11\rangle ) ~.
\end{equation}
The generic structure of a Bell-diagonal state is
$\rho_{BD} = \lambda_1 \vert \psi^-\rangle\langle \psi^- \vert +
             \lambda_2 \vert \phi^-\rangle\langle \phi^- \vert +
             \lambda_3 \vert \phi^+\rangle\langle \phi^+ \vert +
             \lambda_4 \vert \psi^+\rangle\langle \psi^+ \vert$.
With only local unitary operations (so as not to alter the correlations),
all Bell-diagonal states can be transformed to the form
\begin{equation}
\rho_{BD} = \frac{1}{4} \Big( \mathbbm{1} 
          + \sum_{j=1}^{3}r_{j}\sigma_{j}\otimes\sigma_{j} \Big) ~,
\label{rhobd}
\end{equation}
where the real numbers $r_{j}$ are constrained such that all eigenvalues of
$\rho_{BD}$ remain in $[0,1]$. The symmetric form of $\rho_{BD}$ also
implies that it has symmetric discord, i.e., $D_{BD}(B|A) = D_{BD}(A|B)$.

Luo chose the set of measurement bases as $\{V\Pi_{k}^{a}V^{\dagger}\}$,
where $\Pi_{k}^{a}=\vert k\rangle\langle k\vert$ are the projection
operators for the standard basis states ($k=0,1$), and $V$ is an arbitrary
$SU(2)$ rotation matrix. A projective measurement yields the probabilities
$p_{0}=p_{1}=\frac{1}{2}$ and an analytical formula for the classical
correlation,
\begin{equation}
\max_{\{\Pi_{k}^{a}\}} J(A:B) = \left(\frac{1-r}{2}\right)\log_{2}(1-r)
                              + \left(\frac{1+r}{2}\right)\log_{2}(1+r) ~,
\label{jmaxbd}
\end{equation} 
with $r=\max\{|r_{1}|,|r_{2}|,|r_{3}|\}$. 

For the Bell-diagonal states, the reduced density matrices are
$\rho_A = \rho_B = \mathbbm{1}/2$, and the total correlation is
\begin{equation}
I(A:B) = 2+\sum_{i=1}^{4}\lambda_{i}\log_2\lambda_{i} ~,
\label{ibd}
\end{equation}
where the eigenvalues $\lambda_{i}$ of $\rho_{BD}$ are
\begin{eqnarray}
\lambda_{1} &=&(1-r_{1}-r_{2}-r_{3})/4 \nonumber \\ 
\lambda_{2}&=&(1-r_{1}+r_{2}+r_{3})/4 \nonumber \\
\lambda_{3}&=&(1+r_{1}-r_{2}+r_{3})/4 \nonumber \\
\lambda_{4}&=&(1+r_{1}+r_{2}-r_{3})/4. 
\end{eqnarray}
Thus the analytical formula for discord is, as per Eq.~(\ref{disc}),
\begin{eqnarray}
{D}_{BD}(B|A) &=& 2 + \sum_{i=1}^{4}\lambda_{i}\log_2\lambda_{i}
                    - \left(\frac{1-r}{2}\right)\log_{2}(1-r) \nonumber \\
               &&   - \left(\frac{1+r}{2}\right)\log_{2}(1+r) ~.
\label{dbd}
\end{eqnarray}

For a Werner state of the form
\begin{eqnarray}
\rho_{W}(\epsilon) = \frac{1-\epsilon}{4}\mathbbm{1}
                   + \epsilon \vert\psi^- \rangle\langle \psi^- \vert ~,
\label{werner}
\end{eqnarray}
$r_j=-\epsilon$ and $r=\epsilon$. The discord is then given by
\begin{eqnarray}
D_{W}(\epsilon) &=& \frac{1}{4}
  \log_2 \frac{(1-\epsilon)(1+3\epsilon)}{(1+\epsilon)^2} \nonumber \\
                &+& \frac{\epsilon}{4}
  \log_2 \frac{(1+3\epsilon)^3}{(1-\epsilon)(1+\epsilon)^2} \nonumber \\
                &=& \frac{\epsilon^2}{\ln 2} + O(\epsilon^3) ~.
\label{dwerner}
\end{eqnarray}
This expression is plotted versus the purity $\epsilon$ in
Fig.~\ref{werner_corr}, together with the corresponding correlations
$I(A:B)$ and $J(A:B)$.

In practice, the experimental density matrix obtained by tomography
is not necessarily Bell diagonal. We obtain $I(A:B)$ as before, using
Eq.~(\ref{eq5}). To extract the maximum value of $J(A:B)$, we drop the
the off-diagonal terms, keeping only the terms in Eq.~(\ref{rhobd}), and
use Eq.~(\ref{jmaxbd}). In this procedure, discord is overestimated,
whenever the actual direction maximizing $J(A:B)$ is not in the
Bell-diagonal state subspace.

\subsection{Geometric Discord}
Since the maximization of $J(A:B)$ involved in calculating discord is a
hard problem, Dakic {\it et al.} introduced a more easily computable form
of discord based on a geometric measure \cite{Dakic}. For every quantum state there is a set of postmeasurement classical states, and
the geometric discord is defined as the distance between the quantum state
and the nearest classical state,
\begin{equation}
DG(B|A) = \min_{\chi \in \Omega_0}\|\rho-\chi\|^2 ~,
\end{equation}
where $\Omega_0$ represents the set of classical states, and
$\|X-Y\|^2 = {\rm Tr}(X-Y)^2$ is the Hilbert-Schmidt quadratic norm.
Obviously, $DG(B|A)$ is invariant under local unitary transformations.
Analytical formula for computing geometric discord for an arbitrary
$A_{m \times m} \otimes B_{n \times n}$ state of a bipartite quantum
system is available \cite{Luo_geometric}. Recently discovered  ways to
calculate lower bounds on discord for such general states do not require
tomography and, hence, are friendlier experimentally \cite{Rana,Hassan}.

We follow the formalism of Dakic {\it et al.} \cite {Dakic} to obtain
geometric discord for two-qubit states. The two-qubit density matrix
in the Bloch representation is 
\begin{equation}
\rho = \frac{1}{4} \Big( \mathbbm{1} \otimes \mathbbm{1}
     + \sum_{i=1}^{3} x_{i}\sigma_{i}\otimes\mathbbm{1}
     + \sum_{i=1}^{3}y_{i} \mathbbm{1}\otimes\sigma_{i}
     + \sum_{i,j=1}^{3} T_{ij}\sigma_{i}\otimes\sigma_{j} \Big)
\label{bloch}
\end{equation}
where $x_{i}$ and $y_{i}$ represent the Bloch vectors for the two qubits,
and $T_{ij}={\rm Tr}[(\rho(\sigma_{i}\otimes\sigma_{j}))]$ are the components
of the correlation matrix. The geometric discord for such a state is
\begin{equation}
DG(B|A) = \frac{1}{4}(\|x\|^2 + \|T\|^2 - \eta_{\rm max}) ~,
\label{dg}
\end{equation}
where $\|T\|^2 = {\rm Tr}[T^{T}T]$,\ and $\eta_{\rm max}$ is the largest
eigenvalue of $\vec{x}\vec{x}^{T} + TT^{T}$.

\begin{figure}[b]
\hspace*{-0.5cm}
\includegraphics[angle=0,width=7.5cm]{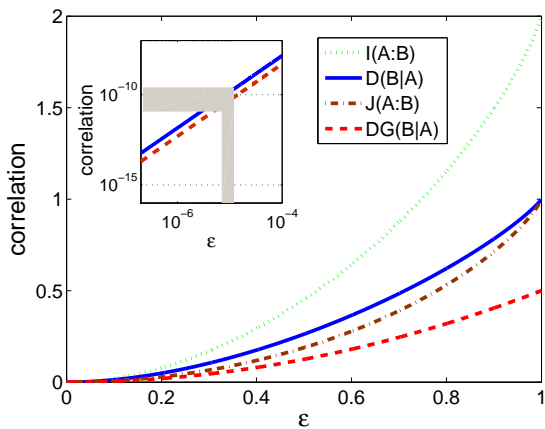}
\caption{Various correlations as functions of the purity $\epsilon$
for Werner states of the form given in Eq.~(\ref{werner}). The inset
shows the range of discord for the purity available in our NMR setup.}
\label{werner_corr}
\end{figure}

For the Werner state, as already mentioned, $x_i = 0 = y_i$ and $T$ is a
diagonal matrix with $T_{ii}=-\epsilon$. Then $\|T\|^2 = 3\epsilon^2$ and
all eigenvalues of $TT^T$ are $\epsilon^2$, yielding
\begin{equation}
DG_{W}(\epsilon) = \frac{1}{4}(3\epsilon^2-\epsilon^2) = \frac{\epsilon^2}{2}.
\end{equation}
This expression is also plotted versus the purity $\epsilon$ in
Fig.~\ref{werner_corr}. Comparison with Eq.~(\ref{dwerner}) reveals that
discord and geometric discord are proportional for low-purity Werner
states. Also, the numerical difference between $D_{W}(\epsilon)$ and
$2 DG_{W}(\epsilon)$ does not exceed 0.027 for all $\epsilon\in[0,1]$.

The Bloch parameters $x_i$, $y_i$, and $T_{ij}$ provide a complete
description of any two-qubit state. So tomographic measurement of these
parameters determines the geometric discord exactly by Eq.~(\ref{dg}).

\section{Experiment}

\begin{figure}
\hspace*{-0.5cm}
\includegraphics[width=8cm]{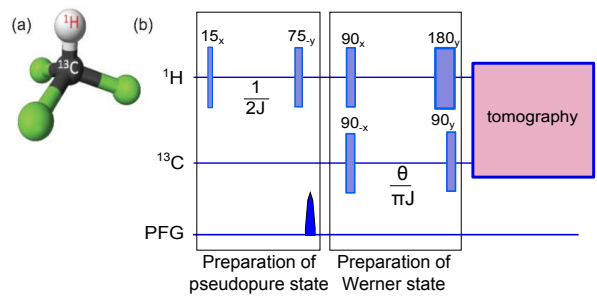}
\caption{(a) The molecular structure of $^{13}$C-chloroform, and
(b) the pulse sequence for discord preparation.
In (b), PFG is the pulse field gradient operation which destroys the
coherences and retains the diagonal elements of the density matrix.}
\label{ps1}
\end{figure}

We now describe experimental evaluation of discord for two-qubit NMR systems.
We measured quantum correlations for two different samples, each forming a
two-qubit system, under different circumstances.\\
Sample 1 is $^1$H and $^{13}$C spins of $^{13}$C-chloroform
[see Fig.~\ref{ps1}a] dissolved in deuterated chloroform (CDCl$_3$).
Both $^1$H and $^{13}$C spins were on-resonant and the scalar coupling
($J$) between the two spins is 219 Hz. For the proton, the $ T_1 $ and $ T_2 $
relaxation time constants are 14.5 and 5.7~s, respectively. For carbon,
they are 21 and 0.25~s, respectively.\\
Sample 2 is $^1$H nuclear pairs of 5-chlorothiophene-2-carbonitrile
(see Fig.~\ref{ps2}b) dissolved in deuterated dymethylsulfoxside (DMSO-D$_6$).
The chemical shift difference ($\Delta\nu$) and scalar coupling ($J$)
between the two spins are 270 and 4.11 Hz, respectively. For each proton,
the $ T_1 $ and $ T_2 $ relaxation time constants are 6.3 and 2.3~s, respectively.

All the experiments were carried out in a Bruker 500-MHz NMR spectrometer
at an ambient temperature of 300 K. Precise radio-frequency (rf) gates for
the experiments were synthesized by numerical optimizations as described in Refs.~\cite{fortunato,maheshsmp}. Decoherence in these systems is mainly due to
the fluctuations in the local magnetic field, caused by random molecular
motions in the presence of intra- and intermolecular dipolar interactions
and chemical shift anisotropies. Traces of paramagnetic impurities also
contribute to spin-relaxation \cite{LevBook}.

The Hamiltonian for a two-qubit NMR system, with spins $I^A$ and $I^B$,
can be written as
\begin{eqnarray}
{\mathscr H} = {\mathscr H}_Z + {\mathscr H}_J ~.
\end{eqnarray}
Here ${\mathscr H}_Z = -\hbar \left( \omega_A I_z^A + \omega_B I_z^B \right)$ is the Zeeman Hamiltonian, characterized by the Larmor frequencies $\omega_A$
and $\omega_B$, and ${\mathscr H}_J = 2\pi J I^A \cdot I^B$ is the indirect
spin-spin coupling Hamiltonian.

In thermal equilibrium at room temperature, $kT$ is much larger than the
Zeeman energy splittings. So the density matrix of a two-qubit system can
be expanded as \cite{cory}
\begin{eqnarray}
\rho_{\rm eq} = \frac{1}{4} e^{-{\mathscr H}/{kT}} \approx
                \frac{1}{4}({\mathbbm 1} + {\overline\rho}_{\rm eq}) ~.
\label{rhoeq}
\end{eqnarray}
The identity ${\mathbbm 1}$ represents a background of uniformly populated
levels, and the traceless part
${\overline\rho}_{\rm eq} = \xi(I_z^A + \frac{\omega_B}{\omega_A}I_z^B)$
is known as the deviation density matrix. Only the traceless part
$\overline\rho$ is manipulated by unitary transformations in all NMR
experiments. For protons in a magnetic field of strength 11.7 T at room
temperature, the small dimensionless number $\xi = \hbar\omega_A / kT
\approx 8 \times 10^{-5}$. The discord for this size of purity is shown in
the inset of Fig.~\ref{werner_corr} by a shaded area.

\subsection{Preparation of non-zero discord states}
We prepared the maximum discord state starting from the zero discord thermal
equilibrium state ($\rho_\mathrm{eq}$) as follows. The pulse sequence in
Fig.~\ref{ps1} was used to prepare discord between $^1$H and $^{13}$C spins
of $^{13}$C-chloroform. For this system, $\omega_C/\omega_H \approx 1/4$.
An initial $\vert 00 \rangle$ pseudopure state was prepared using the spatial
averaging method \cite{cory}, as shown in the first part of Fig.~\ref{ps1}(b).
The transformations of the traceless ${\overline\rho}_{\rm eq}/\xi$ under the
spatial averaging pulse sequence are (as per standard notation \cite{LevBook})
\begin{eqnarray}
&I_z^A + \frac{1}{4}I_z^B& \nonumber \\
&{\downarrow} 15_x^{\circ A}& \nonumber \\
&I_z^A\cos(15^{\circ}) - I_y^A\sin(15^{\circ}) + \frac{1}{4}I_z^B& \nonumber \\
&{\downarrow} \frac{1}{2J} & \nonumber \\
&I_z^A\cos(15^{\circ}) + 2I_x^A I_z^B\sin(15^{\circ}) + \frac{1}{4}I_z^B&
 \nonumber \\
&{\downarrow} 75_{-y}^{\circ A}, G_z& \nonumber \\
&I_z^A\cos15^{\circ}\cos75^{\circ} + 2I_z^A I_z^B\sin15^{\circ}\sin75^{\circ}
 + \frac{1}{4}I_z^B& \nonumber \\
&= \frac{1}{4} (I_z^A + I_z^B + 2I_z^A I_z^B).&
\end{eqnarray}
This pseudopure state was converted into a Werner state, using the
second part of the pulse sequence in Fig.~\ref{ps1}(b) with the delay
$\theta/(\pi J)=1/(2J)$:
\begin{eqnarray}
&\rho_{pp} = \frac{1}{4} \left[ \mathbbm{1} + \frac{\xi}{4}
              \left( I_z^A + I_z^B + 2I_z^A I_z^B \right) \right]& \nonumber \\
 &\downarrow 90_x^{\circ A}, 90_{-x}^{\circ B}& \nonumber\\
 & \frac{1}{4} \left[ \mathbbm{1} + \frac{\xi}{4}
              \left( -I_y^A + I_y^B - 2I_y^A I_y^B \right) \right] &\nonumber \\
 &\downarrow \frac{1}{2J} &\nonumber\\
 & \frac{1}{4} \left[ \mathbbm{1} + \frac{\xi}{4}
              \left( 2I_x^AI_z^B - 2I_z^A I_x^B - 2I_y^A I_y^B \right) \right]& \nonumber \\
 &\downarrow  180_y^{\circ A},90_y^{\circ B} &\nonumber\\
 &\rho_{W} = \frac{1}{4} \left[ \mathbbm{1} + \frac{\xi}{4}
              \left( -2I_x^AI_x^B - 2I_z^A I_z^B - 2I_y^A I_y^B \right) \right] & \nonumber \\
& = \frac{1}{4} \left[ \mathbbm{1} + \frac{\xi}{4}
              \left( - \frac{1}{2}\mathbbm{1} + 2\vert\psi^-\rangle
              \langle\psi^-\vert \right) \right]& \nonumber \\
          &= \frac{1-(\xi/8)}{4}\mathbbm{1}
            + \frac{\xi}{8} \vert\psi^-\rangle\langle\psi^-\vert.&
\label{expwerner}
\end{eqnarray}
Comparing with Eq.~(\ref{werner}), we can see that the relevant purity
parameter in this case is $\epsilon=\xi/8$.

\begin{figure}[b]
\hspace*{-0.2cm}
\includegraphics[angle=0,width=8.3cm]{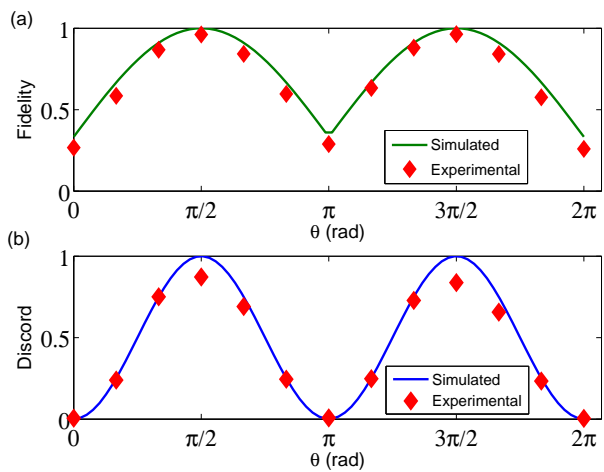}
\caption{(a) Fidelity relative to the Werner state of the
experimental and the simulated states as a function of $\theta$, and
(b) corresponding discord values in units of $\epsilon^2/\ln 2$. The
maximum discord is obtained for the delay parameter $\theta = (2n+1)\pi/2$,
corresponding to preparation of Bell-diagonal states.}
\label{duringgate}
\end{figure} 

We also varied $\theta$ from $0$ to $2\pi$ in 13 steps and, for each
delay, carried out tomography to measure the experimental density matrix
\cite{ChuangPRA99}. The corresponding simulated density matrices are
obtained by assuming perfect pulses and carrying out a calculation similar
to Eq.~(\ref{expwerner}). It can be easily seen that for $\theta$ values
that are odd multiples of $\pi/2$, one obtains Bell-diagonal states.
The fidelity $F$ of a test density matrix $\rho_{\rm test}$
relative to the Werner state is defined as \cite{fortunato}
\begin{equation}
F = \frac{{\rm Tr}[{\overline\rho}_{\rm test} \cdot {\overline\rho}_{W}]}
    {\sqrt{{\rm Tr}[{\overline\rho}_{\rm test}^2]
          ~{\rm Tr}[{\overline\rho}_{W}^2]}} ~.
\label{deffid}
\end{equation}
Fidelities of the experimental and the simulated density matrices,
as functions of $\theta$, are shown in Fig.~\ref{duringgate}(a).
The discord for each value of $\theta$ is obtained using the extensive
measurement method as described in Sec. II-B. Both experimental and
simulated values of the discord are plotted in Fig.~\ref{duringgate}(b),
in units of $\epsilon^2/\ln 2$. The state at $\theta=0$ is related to
the pseudopure $\vert 00 \rangle$ state by local unitary transformations
and, therefore, has zero discord. Otherwise, for $\theta \ne 0$, nonlocal
spin-spin interactions give rise to discord. For $\theta$ equal to odd
multiples of $\pi/2$, one obtains Bell-diagonal states with maximum
discord. For $\theta$ equal to $\pi$, the delay equals the period of the
scalar coupling, implying no transformation, and the discord is periodic
thereafter.

\begin{figure}[b]
\hspace*{-0.5cm}
\includegraphics[angle=0,width=9cm]{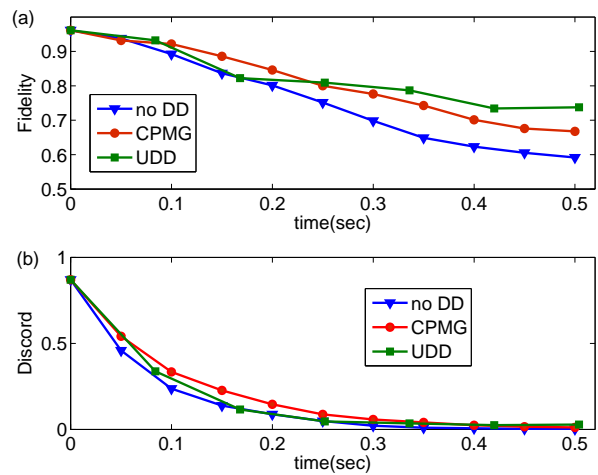}
\caption{(a) Fidelity of the experimental state relative to the Werner
state for various DD schemes, and (b) corresponding discord values in
units of $\epsilon^2/\ln 2$.}
\label{dd}
\end{figure} 

The nonzero discord in the above experiments indicate the intrinsic 
quantumness of nuclear spin systems even at high temperatures. Among
two-qubit states of a given purity, the Werner state possesses the
maximum discord. Next, we investigate the efficiency of dynamical
decoupling schemes in preserving discord.

\subsection{Discord under Dynamical Decoupling}
Dynamical decoupling (DD) is a method of preserving coherences in NMR
by frequent modulation of the spin-environment interaction with the
help of a series of $\pi$ pulses \cite{carr,meiboom,viola99,uhrig}.
The $\pi$-pulses effectively change the sign of the linear spin-bath
interaction, and suitable time spacings between them undo the time
evolution and suppress the $ T_2 $-type decoherence due to the bath.
We applied DD sequences immediately after obtaining the Werner state
as in Eq.~(\ref{expwerner}) and followed that up with tomography.
The CPMG DD involved a series of uniformly spaced $\pi$ pulses separated
by 4-ms delays. For comparison, we label as no-DD the evolution with the
delays but without the pulses. The Uhrig DD (UDD) involved cycles of a
seven pulse sequence \cite{uhrig,soumyaagarwal}. The time-instant $t_j$ of
the $j^{\rm th}$ $\pi$ pulse in each cycle was chosen according to
Uhrig's formula $t_j = 28 \sin^2(\pi j/16)$~ms \cite{uhrig}.

Figure~\ref{dd}(a) shows the time dependence of fidelities relative to the
Werner state for no-DD, a CPMG DD sequence, and UDD sequence. Figure~\ref{dd}(b)
displays the corresponding discord values obtained using the extensive
measurement method. We observe that the DD sequences help in protecting
fidelities of two-qubit quantum states, in agreement with an earlier work
\cite{soumyaagarwal}. On the other hand, there is not much difference
between no-DD and DD schemes in preserving discord. We believe that the
reason is the decay of purity during the DD sequences. This decay is a
$ T_1 $-type decoherence, which the DD sequences do not protect against.
By definition, fidelity looks at the orientation of the quantum state
relative to a target state and not its normalization. On the other hand,
discord depends on the normalization of the state $\overline{\rho}$.
Our experiments indicate that though the DD sequences help prevent the
quantum state from evolving to other quantum states, they are not useful
in keeping the purity from decaying.

\begin{figure}[b]
\includegraphics[width=8cm]{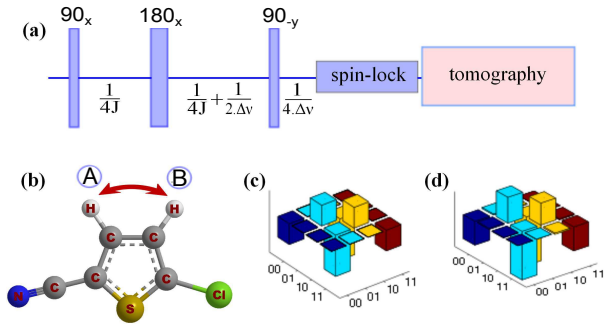}
\caption{The pulse sequence for preparing long-lived singlet states (a),
molecular structure of 5-chlorothiophene-2-carbonitrile (b), and traceless
real parts of the theoretical (c) and the experimental (d) density matrices. 
The experimental Werner state in (d) was obtained with a spin-lock of 16.4~s
and has a fidelity of 0.99.}
\label{ps2}
\end{figure}

\subsection{Discord in Long-lived Singlet States}
Here we considered a pair of nuclear spins of the same isotope, i.e.
{\it Sample 2} shown in Fig.~\ref{ps2}(b). The closeness of their
Larmor frequencies allows a spin-lock procedure \cite{LevJCP09},
where a continuous low-amplitude resonant rf irradiation makes the spins
nutate and drives the spin state toward an isotropic distribution with
all $I_z$ values equally likely. The pulse sequence for the preparation
of a long-lived singlet state is shown in Fig.~\ref{ps2}(a). The rf wave
for the spin-lock had the carrier frequency $(\omega_A+\omega_B)/2$,
the mean Larmor frequency of the two spins, and a nutation frequency
of 2~kHz.

The average Hamiltonian in the interaction frame during the spin-lock interval
is ${\mathscr H}_{\rm int} = 2\pi\hbar J (I_1 \cdot I_2)$, where $I_1$ and
$I_2$ are the spin operators. The singlet state, and the degenerate triplet
states, form an orthonormal eigenbasis of this ${\mathscr H}_{\rm int}$:
\begin{eqnarray}
&& \hspace{-9mm} \vert S_0 \rangle = \frac{1}{\sqrt{2}} (\vert 01\rangle - \vert 10\rangle) ,~
\\
&& \hspace{-9mm} \vert T_1 \rangle = \vert 00\rangle ,~
   \vert T_0 \rangle = \frac{1}{\sqrt{2}} (\vert 01\rangle + \vert 10\rangle) ,~
   \vert T_{-1} \rangle = \vert 11\rangle .
\end{eqnarray}
As described earlier, the NMR system under ordinary conditions exists in a
highly mixed state with a small purity. Leaving out the uniform distribution,
the ground state is the Werner state, which is also called the long-lived
singlet state (LLS) \cite{LevPRL04,LevittJACS04}. The LLS is antisymmetric with
respect to spin exchange, and is not connected to other eigenstates (i.e.
symmetric triplet states) by any symmetry preserving transformations such
as the nonselective rf pulses and the intrapair dipolar interaction.
Therefore, the LLS can survive for durations much longer than other
non-equilibrium spin states. The LLS have been used in NMR experiments for the
study of slow diffusion \cite{BodenJACS05}, for ultraprecise measurement
of scalar interactions \cite{LevPRL09}, for storage and transport of
parahydrogen \cite{GrantJMR08,WarrenScience09}, and for preparation of high
fidelity Bell states and other pseudopure states \cite{soumyasingletini}.

The rf pulses prior to the spin-lock prepare a state which is a mixture of
the $\vert S_0 \rangle$ and $\vert T_0 \rangle$ states,
\begin{equation}
\rho(0) = \frac{1}{4}\mathbbm{1} + \frac{\xi}{4}
( \vert S_0\rangle\langle S_0\vert - \vert T_0\rangle\langle T_0\vert ) ~.
\end{equation}
During the spin-lock, rf irradiation mixes various components of states with
the same spin, and the $\vert T_0 \rangle$ state rapidly equilibrates with
the other triplet states. Furthermore, all other coherences created due to
pulse imperfections also decay towards the background \cite{maheshjmr10}.
On this equilibration, which takes a few seconds, the system reaches the
Werner state,
\begin{equation}
\rho_{\rm LLS} = \frac{1-(\xi/3)}{4}\mathbbm{1}
               + \frac{\xi}{3}\vert S_0 \rangle \langle S_0 \vert
               = \rho_{W}(\epsilon=\xi/3).
\label{rholls}
\end{equation}
This Werner state has a purity that differs from the one prepared from
{\it Sample 1}, i.e., Eq.~(\ref{expwerner}).

To study the evolution of the density matrix state during spin-lock, we
applied the spin-lock for a variable duration $\tau$ and then carried out
tomography to measure the traceless part of the density matrix $\rho(\tau)$.
We find that $\rho(\tau)$ gradually evolves towards the Werner state $\rho_W$,
remains in that state for several tens of seconds, and ultimately decays
towards the uniform state $\mathbbm{1}/4$ that is the asymptotic eigenstate
of the spin-lock evolution after the decay of all spin correlations.

\begin{figure}
\hspace*{-0.5cm} \vspace*{-0.3cm}
\includegraphics[angle=0,width=9cm]{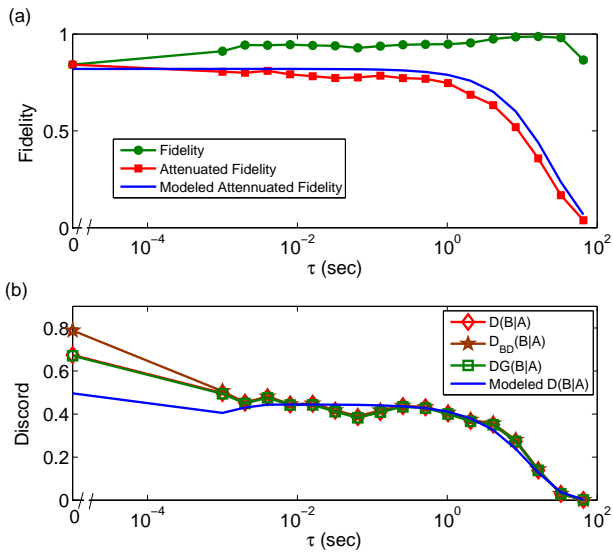}
\caption{(a) Fidelity of the experimental state relative to the Werner
state as a function of the spin-lock duration $\tau$, and
(b) the corresponding values of discord in units of $\epsilon^2/\ln 2$
and geometric discord in units of $\epsilon^2/2$. Discord values were
obtained using the methods described in Sec. II.}
\label{spinlock}
\end{figure}

We monitored fidelity of the experimental state relative to the Werner state
at 17 spin-lock durations, $\tau = 2^n$~ms with $n=\{0,1,\cdots,16\}$).
As shown in Fig.~\ref{spinlock}(a), it starts with a value of 0.85,
reaches a maximum of 0.99 after a few seconds, and then decreases.
The real parts of the deviation density matrices ${\overline\rho}_{W}$ and
${\overline\rho}(\tau)$, with $\tau = 16.4~s$ corresponding to the maximum
fidelity 0.99, are compared in Figs.~\ref{ps2}(c) and \ref{ps2}(d).

Although fidelity is a good measure of how close a test density matrix is
to the target density matrix, it does not capture the global decay of the
purity of the density matrix. To monitor the decay of the purity as well
as the closeness of the traceless parts of the density matrices, we use
attenuated fidelity \cite{fortunato},
\begin{equation}
F_a = \frac{{\rm Tr}[{\overline\rho}(\tau) \cdot {\overline\rho}_{W}]}
      {\sqrt{{\rm Tr}[{\overline\rho}(0)^2]{\rm Tr}[{\overline\rho}_{W}^2]}} ~.
\label{defatnfid}
\end{equation}
It differs from fidelity in normalization, as evident from denominators of
Eq.(\ref{deffid}) and Eq.(\ref{defatnfid}), and decreases as the purity
${\overline\rho}(\tau)$ decays. Figure~\ref{spinlock}(a) also displays
attenuated fidelity as a function of the spin-lock duration. We observe
that it remains close to its initial value 0.85 until about 1~s, and then
drops down. In particular, it starts dropping before the fidelity reaches
its maximum value, and is 0.36 at $\tau = 16.4~s$.

To understand the evolution of the density matrix during spin-lock,
we consider a model consisting of exponential equilibration of the
$\vert T_0\rangle$ state with the other triplet states as well as an
overall exponential decay of the singlet-triplet mixture toward the
uniform identity state:
\begin{equation}
\rho(t) = \frac{1}{4}\mathbbm{1} + e^{-\lambda_2 t} \frac{\xi}{4} 
\Big( \rho_{S0}
    - e^{-\lambda_1 t} \rho_{T0} - (1-e^{-\lambda_1 t})\rho_T \Big) ~.
\label{twoparmodel}
\end{equation}
Here
$\rho_{T} = \frac{1}{3} \left( \vert T_{1} \rangle\langle T_{1} \vert
                             + \vert T_{0} \rangle\langle T_{0} \vert
                             + \vert T_{-1}\rangle\langle T_{-1}\vert \right\}$,
and $\lambda_1$ and $\lambda_2$ are the decay constants. By fitting the
attenuated fidelity of this model to the experimental attenuated fidelity,
as shown in Fig.~\ref{spinlock}(a), we determined $\lambda_1^{-1}=0.75$~ms
and $\lambda_2^{-1}=26$~s.  These values indicate the rapid equilibration 
of the triplet states and the long-lived nature of the singlet state. It
can be noticed that $\lambda_1^{-1}$ is comparable to the rf period during
the spin-lock, and $\lambda_2^{-1}$ is significantly longer than spin-lattice
relaxation time constants ($T_1$ values).

We also measured discord during the spin-lock evolution using the methods
described in Sec. II, and the results are plotted versus the spin-lock
duration in Fig.~\ref{spinlock}(b). The results for the extensive measurement
method $D(B|A)$ and geometric discord $DG(B|A)$ essentially agree (when
scaled by appropriate factors), as expected for accurate methods, and
indicate that measurement errors in our experiments are rather small.
By looking at angular variation of $J(A:B)$ in the extensive measurement
method, at $\tau = 16.4~s$ when the state is closest to the isotropic Werner
state, we estimate that the imperfections in our prepared LLS state give
around 3\% error to discord values.

The discord $D_{BD}(B|A)$ obtained by assuming that the state is of
Bell-diagonal form is an overestimate initially but becomes almost the
same as the other two determinations beyond $\tau = 1$~ms. That means that
artifact off-diagonal coherences are present in our prepared state,
but they decay rapidly on a time scale comparable to $\lambda_1^{-1}$.
The discord value for the two-parameter model of Eq.(\ref{twoparmodel})
is also shown in Fig.~\ref{spinlock}(b). It is accurate once the state
becomes Bell-diagonal, but is unable to model the initial behavior.
The reason for the initial discrepancy is that the off-diagonal components
missing from Eq.(\ref{twoparmodel}) alter both $I(A:B)$ and $J(A:B)$.
Later evolution and the asymptotic vanishing of discord after long
durations of spin-lock is governed by the time scale $\lambda_2^{-1}$.

\section{Conclusions}

Over the years, several hypotheses have been proposed to characterize the
nonclassical part of quantum correlations, e.g., entanglement, violation
of Bell-type inequalities, uncertainty relations, quantum discord, and
so on. They are not all equivalent, especially for mixed states. Quantum
discord aims to be the most inclusive of all of these, and we
have focused our attention to the study of its dynamics in this work.

It is well known that under ordinary NMR conditions, the purity of an
ensemble of nuclear spin systems is below the threshold to exhibit
entanglement. However, successful demonstrations of NMR quantum information
processing indicate that quantum correlations do exist in such ensembles.
In our work, we have, first, revisited the theoretical basis of discord as
well as geometric discord and then have described how they can be produced,
manipulated, and monitored for an experimental density matrix, using the
Werner state as a specific example. We have also implemented several checks
in our investigations to keep track of experimental imperfections.

The experimental study of discord was carried out in two different systems.  
In one system, we studied preparation of discord using an entangling pulse
sequence, and evolution of discord under dynamical decoupling sequences.
Under DD, discord did not show much improvement, but there was clear
improvement in fidelity, which implies that DD protects against $ T_2 $-type
decoherences but not against $ T_1 $-type decoherences. In the second system,
we used the construction of the long-lived singlet state to prepare the Werner
state and examined its evolution under rf spin-lock. We can describe the
accompanying evolution of fidelity and discord reasonably well using a
simple relaxation model. In both systems, the experimentally observed
behavior and the fit parameters are in good agreement with expectations
from theory and simulations.

\section*{Acknowledgments}
The authors are grateful to Anil Kumar and Vikram Athalye for
discussions. S.S. Roy acknowledges support from a CSIR graduate fellowship.
This work was partly supported by the DST project SR/S2/LOP-0017/2009.


\begin{thebibliography}{99}

\bibitem{horodecki}
R. Horodecki {\it et al.}
Rev. Mod. Phys. {\bf 81}, 865 (2009).



\bibitem{Ollivier}
H. Ollivier and W.H. Zurek,
Phys. Rev. Lett. {\bf 88}, 017901 (2002).

\bibitem{vedral}
L. Henderson and V. Vedral,
J. Phys. A {\bf 34}, 6899 (2001).

\bibitem{Braun}
S.L. Braunstein {\it et al.},
Phys. Rev. Lett. {\bf 83}, 1054 (1999).

\bibitem{oliviera}
I.S. Oliveira {\it et al.},
\textit{NMR Quantum Information Processing }(Elsevier, Amsterdam, 2007).

\bibitem{chuang}
L.M.K. Vandersypen {\it et al.},
Nature {\bf 414}, 883 (2001).

\bibitem{Knill}
E. Knill and R. Laflamme,
Phys. Rev. Lett. {\bf 81}, 5672 (1998).

\bibitem{Animesh}
A. Datta, A. Shaji and C.M. Caves,
Phys. Rev. Lett. {\bf 100}, 050502 (2008).

\bibitem{vedral_prx}
K. Modi {\it et al.},
Phys. Rev. X {\bf 1}, 021022 (2011).

\bibitem{vedral_revarticle}
K. Modi {\it et al.},
arXiv:1112.6238 (2011).

\bibitem{Dakic}
B. Dakic, V. Vedral and C. Brukner,
Phys. Rev. Lett. {\bf 105}, 190502 (2010).

\bibitem{serraaug11}
R. Auccaise {\it et al.},
Phys. Rev. Lett. {\bf 107}, 070501 (2011).

\bibitem{laflammeoct11}
G. Passante {\it et al.},
Phys. Rev. A {\bf 84}, 044302 (2011).

\bibitem{soumyaprl}
V. Athalye, S.S. Roy and T.S. Mahesh,
Phys. Rev. Lett. {\bf 107}, 130402 (2011).

\bibitem{white08}
B.P. Lanyon {\it et al.},
Phys. Rev. Lett. {\bf 101}, 200501 (2008).

\bibitem{serradiscordquad}
D.O. Soares-Pinto {\it et al.},
Phys. Rev. A {\bf 81}, 062118 (2010).

\bibitem{footPOVM}
Reference.~\cite{vedral} actually maximizes $J(A:B)$ over generalised POVMs,
and not over orthonormal measurements. It has been shown that rank-1
POVMs suffice for this purpose \cite{animesh_nullity}. In case of
two-dimensional Hilbert spaces that we are dealing with, rank-1
POVMs can be reduced to projective measurements.

\bibitem{animesh_nullity}
A. Datta,
PhD thesis, The University of New Mexico,
arXiv:0807.4490 (2008);
arXiv:1003.5256 (2010).

\bibitem{girolami}
D. Girolami and G. Adesso,
Phys. Rev. A {\bf 83}, 052108 (2011).

\bibitem{chen}
Q. Chen {\it et al.},
Phys. Rev. A {\bf 84}, 042313 (2011).

\bibitem{luo}
S. Luo,
Phys. Rev. A {\bf 77}, 042303 (2008).

\bibitem{Luo_geometric}
S Luo and S Fu,
Phys. Rev. A {\bf 82}, 034302 (2010).

\bibitem{Rana}
S. Rana and P. Parashar,
Phys. Rev. A {\bf 85}, 024102 (2012).

\bibitem{Hassan}
A.S.M. Hassan, B. Lari and P.S. Joag,
Phys. Rev. A {\bf 85}, 024302 (2012)
\bibitem{fortunato}
E.M. Fortunato {\it et al.},
J. Chem. Phys. {\bf 116}, 7599 (2002).

\bibitem{maheshsmp}
T.S. Mahesh and D. Suter, 
Phys. Rev. A {\bf 74}, 062312 (2006).

\bibitem{LevBook}
M.H. Levitt, 
Spin Dynamics: Basics of Nuclear Magnetic Resonance, 2nd edition
(John Wiley and Sons, Chichester, UK, 2008).

\bibitem{cory}
D.G. Cory, M.D. Price and T.F. Havel,
Physica D {\bf 120}, 82 (1998).

\bibitem{ChuangPRA99}
D. Leung {\it et al.},
Phys. Rev. A {\bf 60}, 1924 (1999).

\bibitem{carr}
H.Y. Carr and E.M. Purcell,
Phys. Rev. {\bf 94}, 630 (1954).

\bibitem{meiboom}
S. Meiboom and D. Gill,   
Rev. Sci. Instr. {\bf 29}, 688 (1958).

\bibitem{viola99}
L. Viola, E. Knill and S. Lloyd, 
Phys. Rev. Lett. {\bf 82}, 2417 (1999).

\bibitem{uhrig}
G.S. Uhrig,
Phys. Rev. Lett. {\bf 98}, 100504 (2007).

\bibitem{soumyaagarwal}
S.S. Roy, T.S. Mahesh and G.S. Agarwal,
Phys. Rev. A {\bf 83}, 062326 (2011).

\bibitem{LevJCP09}
G. Pileio and M.H. Levitt,
J. Chem. Phys. {\bf 130}, 214501 (2009). 

\bibitem{LevPRL04}
M. Carravetta, O.G. Johannessen and M.H. Levitt, 
Phys. Rev. Lett. {\bf 92}, 153003 (2004). 

\bibitem{LevittJACS04}
M. Carravetta and M.H. Levitt, 
J. Am. Chem. Soc. {\bf 126}, 6228 (2004).

\bibitem{BodenJACS05}
S. Cavadini {\it et al.}
J. Am. Chem. Soc. {\bf 127}, 15744 (2005).

\bibitem{LevPRL09}
G. Pileio, M. Carravetta and M.H. Levitt,
Phys. Rev. Lett. {\bf 103}, 083002 (2009). 

\bibitem{GrantJMR08}
A.K. Grant and E. Vinogradov, 
J. Magn. Reson. {\bf 193}, 177 (2008).

\bibitem{WarrenScience09}
W.S. Warren {\it et al.},
Science {\bf 323}, 1711 (2009).

\bibitem{soumyasingletini}
S.S. Roy and T.S. Mahesh,
Phys. Rev. A {\bf 82}, 052302 (2010).

\bibitem{maheshjmr10}
S.S. Roy and T.S. Mahesh,
J. Magn. Reson. {\bf 206}, 127 (2010).

\end{thebibliography}
\end{document}